\documentclass[aps,pre,preprint,superscriptaddress,showpacs,amssymb,floatfix]{revtex4}
\usepackage{graphicx}
\usepackage{graphics}
\usepackage{amsmath}
\usepackage{tabularx}
\usepackage{color}
\usepackage{doi}
\usepackage{textcomp}
\newcolumntype{Y}{>{\centering\arraybackslash}X}

\begin{document}

\title{Scaling corrections of Majority Vote model on Barabasi-Albert networks}

\author{T. F. A. Alves}
\affiliation{Departamento de F\'{\i}sica, Universidade Federal do Piau\'{i}, 57072-970, Teresina - PI, Brazil}
\author{G. A. Alves}
\affiliation{Departamento de F\'{i}sica, Universidade Estadual do Piau\'{i}, 64002-150, Teresina - PI, Brazil}
\author{F. W. S. Lima}
\affiliation{Departamento de F\'{\i}sica, Universidade Federal do Piau\'{i}, 57072-970, Teresina - PI, Brazil}
\author{A. Macedo-Filho}
\affiliation{Campus Prof.\ Antonio Geovanne Alves de Sousa, Universidade Estadual do Piau\'i, 64260-000, Piripiri - PI, Brazil}

\date{Received: date / Revised version: date}

\begin{abstract}

We consider two consensus formation models coupled to Barabasi-Albert networks, namely the Majority Vote model and Biswas-Chatterjee-Sen model. Recent works point to a non-universal behavior of the Majority Vote model, where the critical exponents have a dependence on the connectivity while the effective dimension $D_\mathrm{eff} = 2\beta/\nu + \gamma/\nu$ of the lattice is unity. We considered a generalization of the scaling relations in order to include logarithmic corrections. We obtained the leading critical exponent ratios $1/\nu$, $\beta/\nu$, and $\gamma/\nu$ by finite-size scaling data collapses, as well as the logarithmic correction pseudo-exponents $\widehat{\lambda}$, $\widehat{\beta}+\beta\widehat{\lambda}$, and $\widehat{\gamma}-\gamma\widehat{\lambda}$. By comparing the scaling behaviors of the Majority Vote and Biswas-Chatterjee-Sen models, we argue that the exponents of Majority Vote model, in fact, are universal. Therefore, they do not depend on network connectivity. In addition, the critical exponents and the universality class are the same of Biswas-Chatterjee-Sen model, as seen for periodic and random graphs. However, the Majority Vote model has logarithmic corrections on its scaling properties, while Biswas-Chatterjee-Sen model follows usual scaling relations without logarithmic corrections.

\end{abstract}


\maketitle

\section{Introduction}

Barabasi-Albert (BA) model is the toy model of scale-free networks\cite{Barabasi-1999, Albert-2002, Newman-2002, Dorogovtsev-2003, Barrat-2004, Boccaletti-2006, Barrat-2008, Cohen-2010, Barabasi-2016}. Scale-free networks are important because they are ubiquitous in nature and society. Many real networks are known to be scale-free, for example, the network of human sexual contacts\cite{Liljeros-2001}, the world wide web\cite{Barabasi-1999, Dorogovtsev-2003, Barabasi-2016}, the transport network\cite{Barrat-2004}, the citation network\cite{Price-1965, Redner-1998}, the network of scientific collaborations\cite{Newman-2001, Barabasi-2002} among many others. A fundamental feature of a scale-free network is the presence of hubs, which are the highly connected nodes\cite{Cohen-2010, Barabasi-2016}. If hubs are present, we can expect a change in the system behavior\cite{Cohen-2010, Barabasi-2016}. Other fundamental properties are the ultra small-world property, and the presence of degree correlations\cite{Cohen-2010, Barabasi-2016}. 

Scale-free networks are a special case of power-law degree distribution networks, following
\begin{equation}
P(k) \sim k^{-\lambda}.
\end{equation}
where $k$ is the number of connected neighbors, i.e., the degree, $\lambda$ is the power-law exponent, and $P$ is the probability distribution. Regarding its average degree $\left<k\right>$ and second moment $\left<k^2\right>$, power-law networks can be sorted as
\begin{itemize}
  \item Power-law networks with $\lambda \leq 2$: All distribution momenta diverge, even the average degree $\left<k\right>$;
  \item Power-law networks with $2 < \lambda \leq 3$: Average degree $\left<k\right>$ is finite and all other momenta diverge;
  \item Power-law networks with $\lambda > 3$: Average degree $\left<k\right>$ and second moment $\left<k^2\right>$ are finite;
\end{itemize}
Power-law networks with $2 < \lambda \leq 3$ lacking scale in the sense of unbound fluctuations on the average degree\cite{Cohen-2010, Barabasi-2016}, hence the name scale-free. BA networks have $\lambda=3$, therefore, they are on a marginal situation where networks behave as random graphs with bounded degree fluctuations and scale-free graphs with unbounded degree fluctuations. In fact, for BA networks, $\left<k^2\right>$ diverges logarithmically as
\begin{equation}
\frac{\left<k^2\right>}{\left<k\right>} \sim \frac{z}{2}\ln N,
\label{banetwork-fluctuations}
\end{equation}
where $z$ is the number of bonds a newly added node will have when inserted into the growing network, i.e., the connectivity.

Unbounded degree fluctuations introduce non-trivial effects on phase transitions\cite{PhysRevLett.96.038701, PhysRevLett.98.258701, RevModPhys.80.1275}. One well studied example is the Contact Process (CP) model on a special class of uncorrelated networks: the Uncorrelated Configuration Model (UCM)\cite{RevModPhys.80.1275, PhysRevLett.98.258701}. The UCM is an algorithm to generate uncorrelated scale-free networks with an externally controlled power-law exponent $\lambda$\cite{PhysRevE.71.027103}. Publication of Heterogeneous Mean Field (HMF) theory for scale-free networks was followed by an intense debate if the critical behavior of the CP model on UCM networks obeys HMF theory\cite{PhysRevLett.96.038701, PhysRevLett.98.258701, PhysRevE.84.066102}, settled by the fact that the critical behavior of the CP model on UCM networks is subjected to scaling corrections.

Considering the special case of UCM networks with $\lambda=3$, HMF theory predicts logarithmic corrections to scaling\cite{PhysRevE.84.066102}. In the same way, results from a special Mean Field theory, applied on BA networks, predict an extra logarithmic dependence in the critical behavior of the CP model order parameter\cite{PhysRevE.86.026117}. Furthermore, there are some other examples of equilibrium and non-equilibrium models whose critical behavior is subjected to logarithmic corrections\cite{Salas1997, kenna:2012, PhysRevLett.96.115701, PhysRevLett.97.155702, PhysRevE.82.011145, jstat.2017.123302}. 

A recent work\cite{2019arXiv190504595V} states that there is a non-universal behavior in a particular consensus formation model\cite{Galam2008}, called Majority Vote (MV) model\cite{MJOliveira1992, Pereira2005, Yu2017, Wu2009, Crochik2005, Vilela2009, Lima2012, Vieira2016, Krawiecki2018, Stanley2018, Alves_2019} on BA networks. Ref. \cite{2019arXiv190504595V} considered a modified version of the MV model where the individuals can have three discrete opinions and its results pointed to varying $1/\nu$, $\beta/\nu$ and $\gamma/\nu$ exponent ratios when changing $z$, but maintaining the effective dimension $D_{\mathrm{eff}}$, defined as
\begin{equation}
D_\mathrm{eff} = 2\beta/\nu + \gamma/\nu,
\label{effectivedimension}
\end{equation}
equal to unity. The same non-universal behavior is reported in Ref.\cite{Lima_2006}, on the usual MV model with two opinion states with $Z_2$ symmetry. Motivated by these two works, we revisit the simpler version of the two-state MV model on BA networks to investigate if this non-universal behavior with varying exponent ratios are, in fact, universal, when properly including scaling corrections as seen for the CP model. We believe the unbounded degree fluctuations can introduce logarithmic corrections in the MV model critical behavior.

In addition, we considered Biswas-Chatterjee-Sen (BCS) model, inspired by wealth exchanges in an open market\cite{BISWAS20123257, PhysRevE.94.062317}. The BCS model has a continuous phase transition in the same universality class of Ising and the MV models on periodic lattices\cite{BISWAS20123257, PhysRevE.94.062317}, however, differently from the MV model, the BCS model can accomplish for individuals assuming an interval of continuous opinion states. Consensus depends on two parameters: a conviction parameter and an affinity parameter, however, for the sake of simplicity, we included only the affinity dependence on the dynamics. In a round of the dynamics, a random bond of the network is randomly selected and the two neighbors can influence one to the other. The reason we considered this model is that we expect that the critical behavior of the BCS model will follow usual scaling relations without logarithmic corrections. Indeed, as we will see, updated states of the dynamics depend only on a pair of nodes (not all neighbors), turning the dynamics insensible to the unbounded degree fluctuations.

In summary, our main objective is comparing the critical behavior of the MV and BCS models. We argue that these models still fall in the same universality class for BA networks, however, we should include logarithmic corrections to the scaling of the MV model. This paper is organized as follows: in section II we describe the MV and BCS models and the finite-size scaling relations with logarithmic corrections, in section III we discuss our numerical results and in section IV we present our conclusions.

\section{Models and Scaling}

\subsection*{Barabasi-Albert Networks}

\begin{figure}[h]
\begin{center}
\includegraphics[scale=0.1]{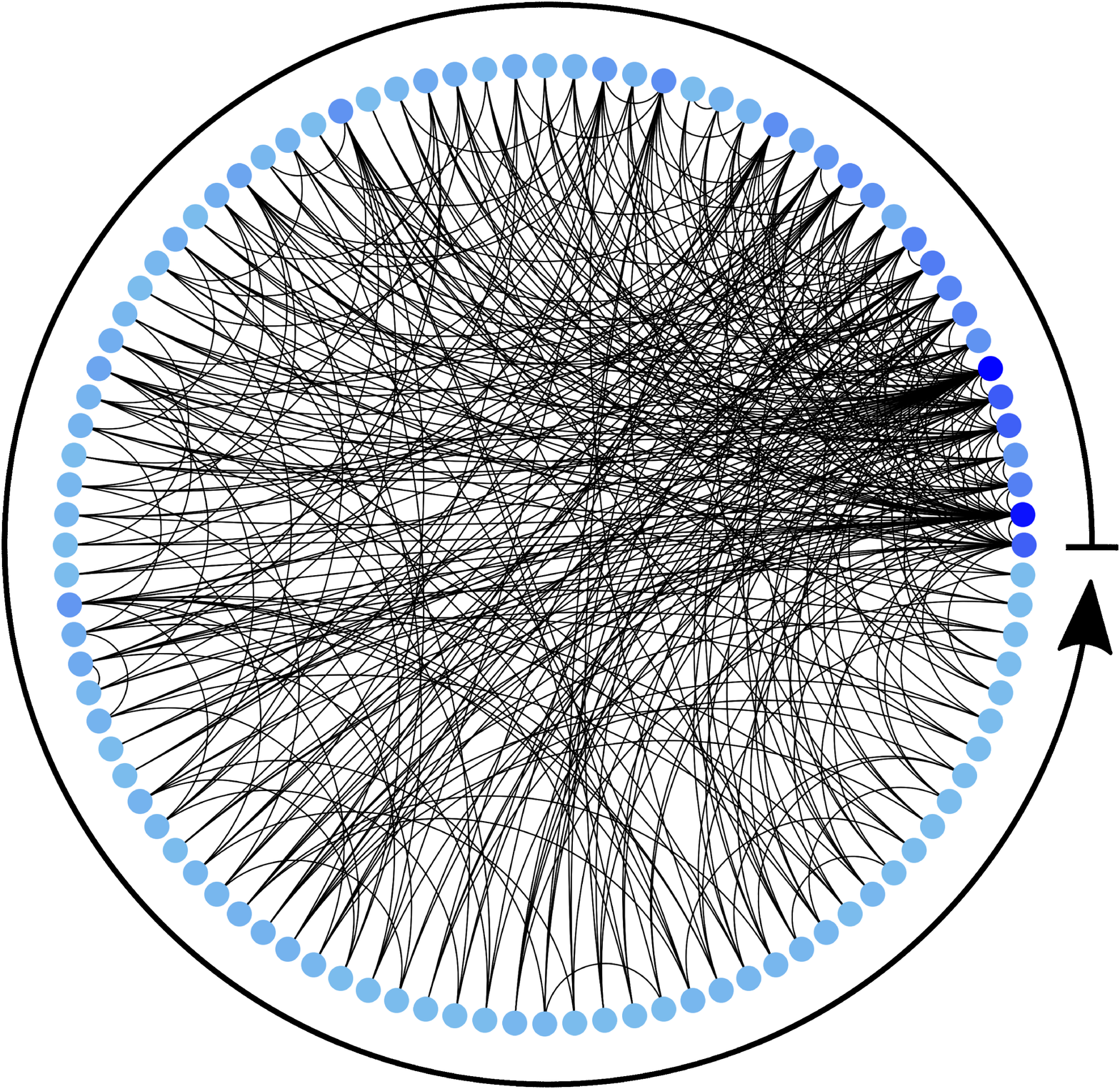} 
\end{center}
\caption{(Color Online) Random realization of a Barabasi-Albert network with 100 nodes, arranged in a circumference. Older nodes in the growing process are placed in smaller angles, measured from the right horizontal direction and nodes with greater degrees are colored in darker shades of grey (blue). Curves represent bonds between nodes. Note the presence of hubs (highly connected nodes) and the first player advantage (older nodes are likely to become hubs).}
\label{banetwork}
\end{figure}

We consider in this work, two models of consensus formation, where both models are coupled to BA networks. We begin by discussing the building algorithm of BA networks\cite{Cohen-2010, Barabasi-2016}. To build BA networks with $N$ nodes, we should start from a complete graph with $z<N$ nodes, and then, grow the graph until it has $N$ nodes by adding one node at a time. Every newly added node starts with $z$ bonds, connecting the newly added node with randomly chosen $z$ already added nodes, according to the preferential attachment probability. Preferential attachment means that the probability $P_i$ of a new node attaching with an older node $i$ is proportional to its degree $k_i$, i.e.,
\begin{equation}
   P_i(k_i) = \frac{k_i}{\sum_j k_j}.
\end{equation}
Growing and preferential attachment are some of the mechanisms that originate network hubs. We show, in Fig.(\ref{banetwork}), a random realization of a BA network with 100 nodes. Note the presence of hubs and the first player advantage: older nodes are likely to become hubs.

\subsection*{The MV and BCS Models}

In the following, we present the first considered model in this work, namely the two-state MV model\cite{MJOliveira1992, Pereira2005, Yu2017}. The MV model dynamics has the following rules:
\begin{enumerate}
   \item For each node of the network, we assign one spin variable $s_{i} = \pm 1$, corresponding to two opinion states. We start the dynamics by randomly selecting the opinion state for each node;
   \item At each time step, we randomly choose one node $j$ to be updated;
   \item Then, we try a spin flip with a probability $\omega(s_{j})$, written as
\begin{equation}
   \omega(s_{j}) = \frac{1}{2} \left[1-(1-2q)s_{j}S\left(\sum_{\delta=1}^{z_{j}} s_{\delta} \right)\right],
\label{mvfliprate}
\end{equation}   
where the index summation $\delta$ runs over all $z_j$ nearest neighbors of the $j-$th vertex and $S(x)$ is the \textit{signal} function, associated with the neighborhood majority opinion
\begin{equation}
S(x) = \left\lbrace \begin{array}{cl}
-1, & \mathrm{if\ }x<0; \\
0,  & \mathrm{if\ }x=0; \\
1,  & \mathrm{if\ }x>0.
\end{array}
\right.
\end{equation}
One should note that in case of no local majority, the $j$-th node can assume any opinion state with $\omega=1/2$. The noise parameter $q$ induces a continuous phase transition from a consensus phase to a no-consensus phase, analogous to the ferro-paramagnetic phase transition.
\end{enumerate}

The second considered model in this work is the BCS model\cite{BISWAS20123257, PhysRevE.94.062317}. Its dynamics has the following rules:
\begin{enumerate}
   \item For each node of the network is assigned one continuous opinion variable $s_i \equiv o_{i}$ in the $[-1,1]$ interval. We start the dynamics by randomly selecting the opinion state by each network node by generating $N$ uniform float random numbers in the $[-1,1]$ interval.
   \item At each time step, we randomly choose one node of the network to be updated;
   \item Then, we randomly select one of its bonds and set the affinity $\mu_{i,j}$ of the bond. The affinity parameter is an annealed random uniform variable in the interval $[0,1]$ which can be turned negative with a probability $q$. The noise parameter $q$ for the BCS model acts in an analogous way of the MV model;
   \item The two nodes $i$ and $j$ will be updated according to the following expressions
\begin{eqnarray}
   o_i(t+1) &=& o_i(t) + \mu_{i,j} o_j(t), \nonumber \\
   o_j(t+1) &=& o_j(t) + \mu_{i,j} o_i(t),
\end{eqnarray}
where $o_i(t)$ and $o_j(t)$ are the older opinion states, and $o_i(t+1)$ and $o_j(t+1)$ are the updated opinion states;
   \item If any of the updated states $o_{i,j}(t+1)$ of nodes $i$ and $j$ satisfy $o_{i,j}(t+1)>1$, they will assume the value $o_{i,j}(t+1)=1$ to maintain the opinion states bounded on $[-1,1]$ interval. Analogously, if $o_{i,j}(t+1)<-1$, the updated states will become $o_{i,j}(t+1)=-1$. This introduces non-linearity to the model.
\end{enumerate}

One should note the main difference between these models regarding interactions between any node and its connected neighbors. Opinion states of the MV model are updated according to opinion states of all of its neighbors in a way any node can follow the neighboring majority opinion with probability $q$, otherwise, it will follow the converse with probability $1-q$. Meanwhile, for the BCS model, updates of any node depend on only one randomly selected neighbor. This main difference can account for different scaling behaviors of the MV and BCS models. In fact, the BCS model will follow usual scaling on BA networks, while scaling of the MV model should be prone to logarithmic corrections.

\subsection*{Observables and critical behavior}

After describing the considered models, we present the needed observables to identify their critical behavior. The main observable is the opinion balance $m$, analogous to the magnetization of magnetic equilibrium systems
\begin{equation}
m = \left\vert \frac{1}{N} \sum_i s_i \right\vert.
\end{equation}
From the opinion balance, one can calculate the order parameter by averaging $m$. BA networks are random networks, so one should do an ensemble average, on the time series resulting from the dynamical evolution, i.e. from a Monte Carlo Markov Chain (MCMC), and a quenched average, done on random realizations of the network. For each random realization, one should evolve dynamics to a stationary state, and then collect an ensemble composed of a temporal series. The order parameter $M$, its respective susceptibility $\chi$, and Binder's fourth-order cumulant $U$ are given by the following relations, respectively\cite{MJOliveira1992}
\begin{eqnarray}
M(q)&=& \left[ \langle m \rangle \right], \nonumber \\
\chi(q) &=& \left[ N (\langle m^{2} \rangle - \langle m \rangle ^{2}) \right], \nonumber \\
U(q) &=& \left[ 1 - \frac{\langle m^{4} \rangle}{3\langle m^{2} \rangle^{2}} \right], \label{observables}
\end{eqnarray}
where the symbol $\langle ... \rangle$ represents the average of a time series and the symbol $\left[ ... \right]$ represents the quench average. All observables are functions of the noise parameter $q$. 

We conjecture that the observables written in Eq.(\ref{observables}) should obey the following finite-size scaling (FSS) relations
\begin{eqnarray}
M &=& N^{-\beta/\nu} \left( \ln N \right)^{\widehat{\beta}+\beta\widehat{\lambda}} f_{M}\left( N^{1/\nu} \left( \ln N \right)^{-\widehat{\lambda}} \left(q-q_{c}\right) \right), \nonumber \\
\chi &=& N^{\gamma/\nu} \left( \ln N \right)^{\widehat{\gamma}-\gamma\widehat{\lambda}} f_{\chi}\left( N^{1/\nu} \left( \ln N \right)^{-\widehat{\lambda}} \left(q-q_{c}\right) \right), \nonumber \\
U &=& f_{U}\left( N^{1/\nu} \left( \ln N \right)^{-\widehat{\lambda}} \left(q-q_{c}\right) \right),\label{observables-fss}
\end{eqnarray}
respectively, where $1/\nu$, $\beta/\nu$, and $\gamma/\nu$ are the critical exponent ratios, $\widehat{\lambda}$, $\widehat{\beta}+\beta\widehat{\lambda}$, and $\widehat{\gamma}-\gamma\widehat{\lambda}$ are the scaling correction pseudo-exponents, $q_c$ is the critical noise and $f_{M,\chi,U}$ are the finite-size scaling functions. Note that if $\widehat{\lambda}=0$, $\widehat{\beta}=0$, and $\widehat{\gamma}=0$, usual scaling relations are recovered. Pseudo-exponents obey the following scaling relation\cite{kenna:2012, PhysRevLett.96.115701, PhysRevLett.97.155702, PhysRevE.82.011145}
\begin{equation}
2\widehat{\beta}-\widehat{\gamma} = -d\nu\widehat{\lambda},
\label{pseudoexponent-scalingid}
\end{equation}
and by combining the scaling relation written in Eq. (\ref{pseudoexponent-scalingid}) with $d=D_{\mathrm{eff}}$ where $D_{\mathrm{eff}}$ is given in Eq.(\ref{effectivedimension}), one can obtain
\begin{equation}
\widehat{\gamma}-\gamma\widehat{\lambda} = 2\left(\widehat{\beta}+\beta\widehat{\lambda}\right).
\label{pseudoexponent-scalingid-2}
\end{equation}

To obtain the relevant observables, we performed MCMC's on BA networks with sizes $N=2500$, $N=3600$, $N=4900$, $N=6400$, $N=8100$, and $N=10000$ where $N$ is the network size. In addition, we considered different connectivities to investigate the non-universal behavior\cite{2019arXiv190504595V,Lima_2006}. For each size and connectivity, we simulated $128$ random network realizations to make quench averages. For each network replica, we considered $10^5$ MCMC steps to let the system evolve to a stationary state and another $10^5$ MCMC steps to collect $10^5$ values of the opinion balance to measure the observables. One MCMC step for the MV model is defined as the update of $N$ spins, while a MCMC step for the BCS model is defined as the update of $N$ node pairs, connected by a bond. Statistical errors were calculated by using the ``jackknife'' resampling technique\cite{Tukey1958}. 

\section{Results and Discussion}

\begin{figure}[p]
\begin{center}
\includegraphics[scale=0.16]{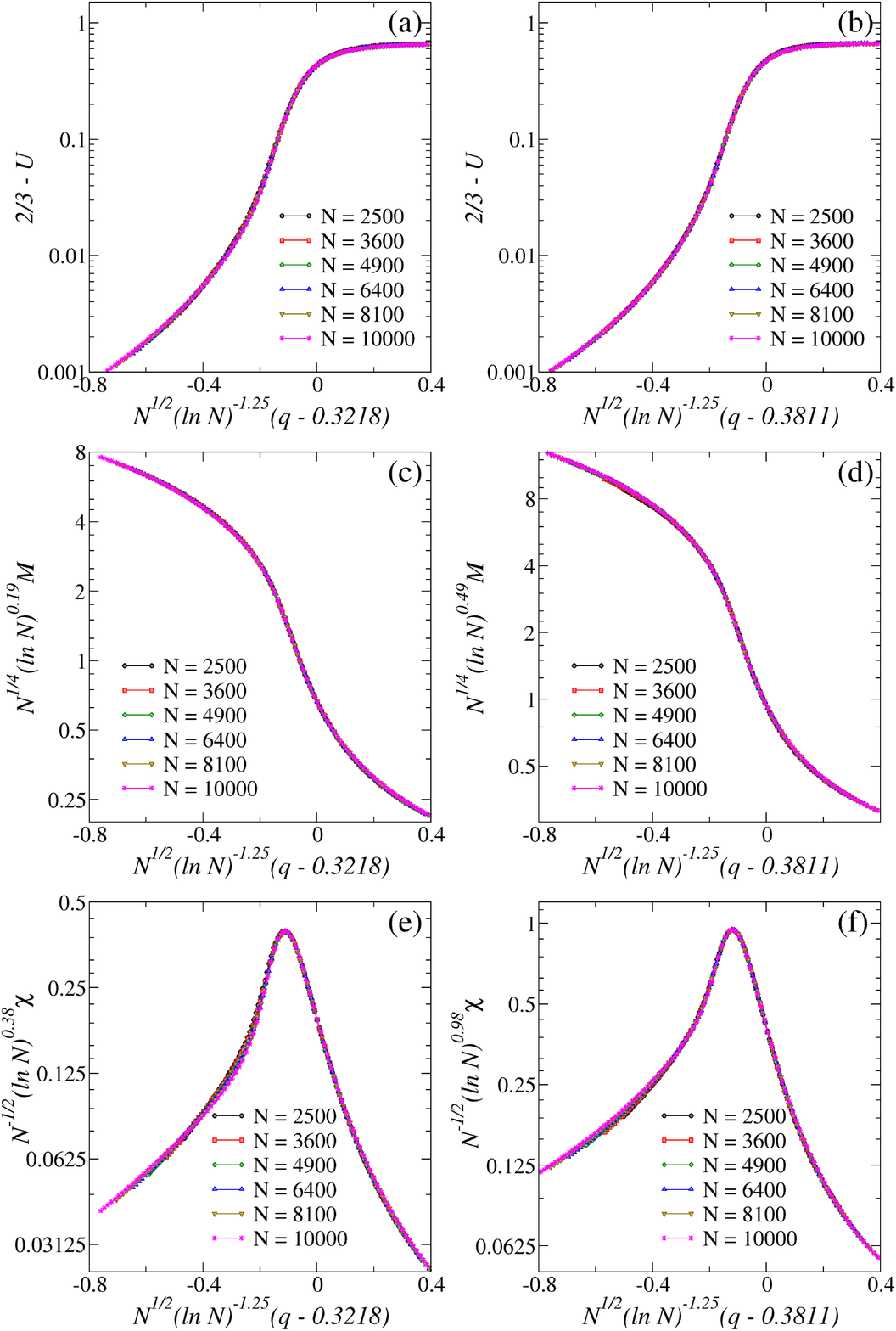} 
\end{center}
\caption{(Color Online) Scaled results for the MV model on BA networks according to FSS relations (\ref{observables-fss}). In panels (a), (c), and (e), we show our numerical data for Binder cumulant $U$, the order parameter $M$, and susceptibility $\chi$ for $z=4$. The same is shown in panels (b), (d), and (f) for $z=8$. We see that our numerical results suggest that the leading critical exponents of BA networks are the same of Erd\"os-Renyi random graphs\cite{Pereira2005}, and they are independent of connectivity $z$. Statistical errors are smaller than symbols.}
\label{mv-collapsed-results}
\end{figure}

We show our numerical results for the MV model on BA networks in Fig.(\ref{mv-collapsed-results}), collapsed by using logarithmic corrected scaling relations written on Eq. (\ref{observables-fss}), combined with the values of the leading critical exponents in Tab.(\ref{criticalbehaviortable}). All numerical results are consistent with a continuous phase transition with critical noises $q_c$ depending on the connectivity. Numerical values of critical noises and logarithmic correction pseudo-exponents used in data collapses for the MV model are presented in Tab.(\ref{criticalbehaviormvtable}).

\begin{table}[h]
\begin{center}
\begin{tabularx}{0.5\textwidth}{YY}
\hline
Critical exponent ratios & Exact values \\
\hline
$1/\nu$      & $1/2$  \\
$\beta/\nu$  & $1/4$  \\
$\gamma/\nu$ & $1/2$  \\
\hline                 
\end{tabularx}
\end{center}
\caption{Leading Mean Field critical exponents for both MV and BCS models.}
\label{criticalbehaviortable}
\end{table}

\begin{table}[h]
\begin{center}
\begin{tabularx}{\textwidth}{Y|YYYY}
\hline
Connectivity & $q_c$ & $\widehat{\lambda}$ & $\widehat{\beta}+\beta\widehat{\lambda}$ & $\widehat{\gamma}-\gamma\widehat{\lambda}$ \\
\hline
$z=4$  & $0.3218$ & $1.25$ & $0.19$ & $0.38$ \\
$z=5$  & $0.3460$ & $1.25$ & $0.31$ & $0.62$ \\
$z=6$  & $0.3605$ & $1.25$ & $0.40$ & $0.80$ \\
$z=7$  & $0.3727$ & $1.25$ & $0.46$ & $0.92$ \\
$z=8$  & $0.3811$ & $1.25$ & $0.49$ & $0.98$ \\
$z=9$  & $0.3889$ & $1.25$ & $0.52$ & $1.05$ \\
$z=10$ & $0.3953$ & $1.25$ & $0.55$ & $1.10$ \\
$z=15$ & $0.4160$ & $1.25$ & $0.60$ & $1.20$ \\
$z=20$ & $0.4273$ & $1.25$ & $0.60$ & $1.20$ \\
\hline                 
\end{tabularx}
\end{center}
\caption{All estimated critical noises and scaling correction pseudo-exponents for the MV model. Note that the pseudo-exponents satisfy the scaling relation written on Eq.(\ref{pseudoexponent-scalingid}) with $d=D_{\mathrm{eff}}$, where $D_{\mathrm{eff}}$ is given in Eq.(\ref{effectivedimension}) and are functions of connectivity $z$ for the MV model. Meanwhile, the BCS model follows clean scaling behavior without logarithmic corrections.}
\label{criticalbehaviormvtable}
\end{table}

In addition, we used in data collapses showed in Fig.(\ref{mv-collapsed-results}), the same leading critical exponents and effective dimension $D_{\mathrm{eff}}$ of the MV model on Erd\"{o}s-Renyi random graphs\cite{Pereira2005}. The leading critical exponents are summarized in Tab.(\ref{criticalbehaviortable}), in a way that the MV model on BA networks has the same critical behavior on random graphs, but with the introduction of logarithmic corrections. Logarithmic corrections are the main consequence of a marginal criticality. In fact, BA networks, with logarithmic diverging degree fluctuations are in the frontier between random graphs and scale-free networks.

From our numerical results of the MV model, we see that the scaling of the observables obeys the same critical leading exponent ratios presented in Tab.(\ref{criticalbehaviortable}), irrespective of its connectivity $z$. Indeed, our results suggest a universal behavior on BA networks when considering scaling corrections. Note that the estimated values of pseudo-exponents, summarized in Tab.(\ref{criticalbehaviormvtable}) are functions of the connectivity $z$ with the exception of $\hat{\lambda}$ pseudo-exponent. However, our numerical collapses do not rule out a slightly varying $\hat{\lambda}$ because values in the interval $[1.25,1.45]$ give equally good data collapses.

\begin{figure}[h]
\begin{center}
\includegraphics[scale=0.3]{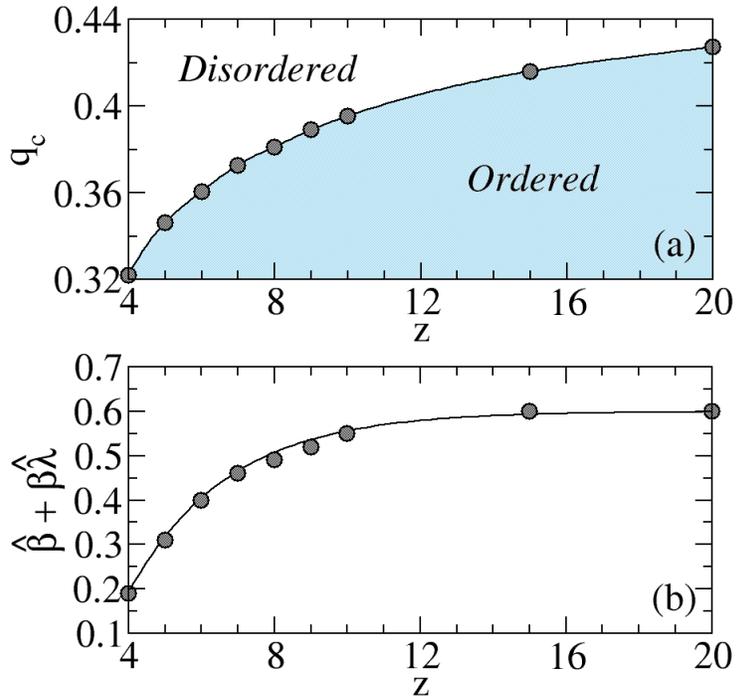} 
\end{center}
\caption{(Color Online) In panel (a) we show the critical noises of the MV model as functions of the connectivity $z$ in circles. Critical noises saturate at $q_c=0.5$ when increasing the connectivity. In the grey (blue) region, we have the ordered phase where a consensus state is reached and in the white region, we have the disordered phase. In panel (b) we show the logarithmic correction pseudo exponent $\widehat{\beta}+\beta\widehat{\lambda}$ as function of the connectivity $z$ in circles. The same saturation pattern when increasing the connectivity is seen from pseudo-exponent data. In both panels, the curve is only a guide to the eye.}
\label{saturation-noises-and-pseudoexponents}
\end{figure}

In consequence, if one does not use the generalized scaling relations, apparent leading exponent ratios should depend on $z$ because of increasing strength of diverging fluctuations in finite networks, as seen from Eq.(\ref{banetwork-fluctuations}). Our data can be collapsed with apparent critical exponent ratios. This is consistent with varying exponent ratios reported in previous works\cite{2019arXiv190504595V, Lima_2006}. We see a clear saturation pattern on the critical noises in Fig.(\ref{saturation-noises-and-pseudoexponents}), and the saturation value is the limiting value of $q=0.5$ for the complete graph\cite{PhysRevE.96.012304}. The same saturation pattern is seen on the critical pseudo-exponents for $z>10$, shown in Fig.(\ref{saturation-noises-and-pseudoexponents}). 

In addition, we should stress about a feature of the data collapses for the MV model. We noted it was easier to collapse network sizes ranging from $N=2500$ to $N=10000$ for greater values of connectivity $z$, meanwhile, fitting the curves with $N=2500$ and $N=3600$ was more difficult for connectivities $z \leq 5$. This is linked with the fact of capturing the degree distribution for lower connectivities needs more time (and nodes) in the growing process of BA scale-free networks. Indeed, for lower connectivities, we need to simulate the MV model on bigger networks to properly capture its critical behavior.

\begin{figure}[p]
\begin{center}
\includegraphics[scale=0.16]{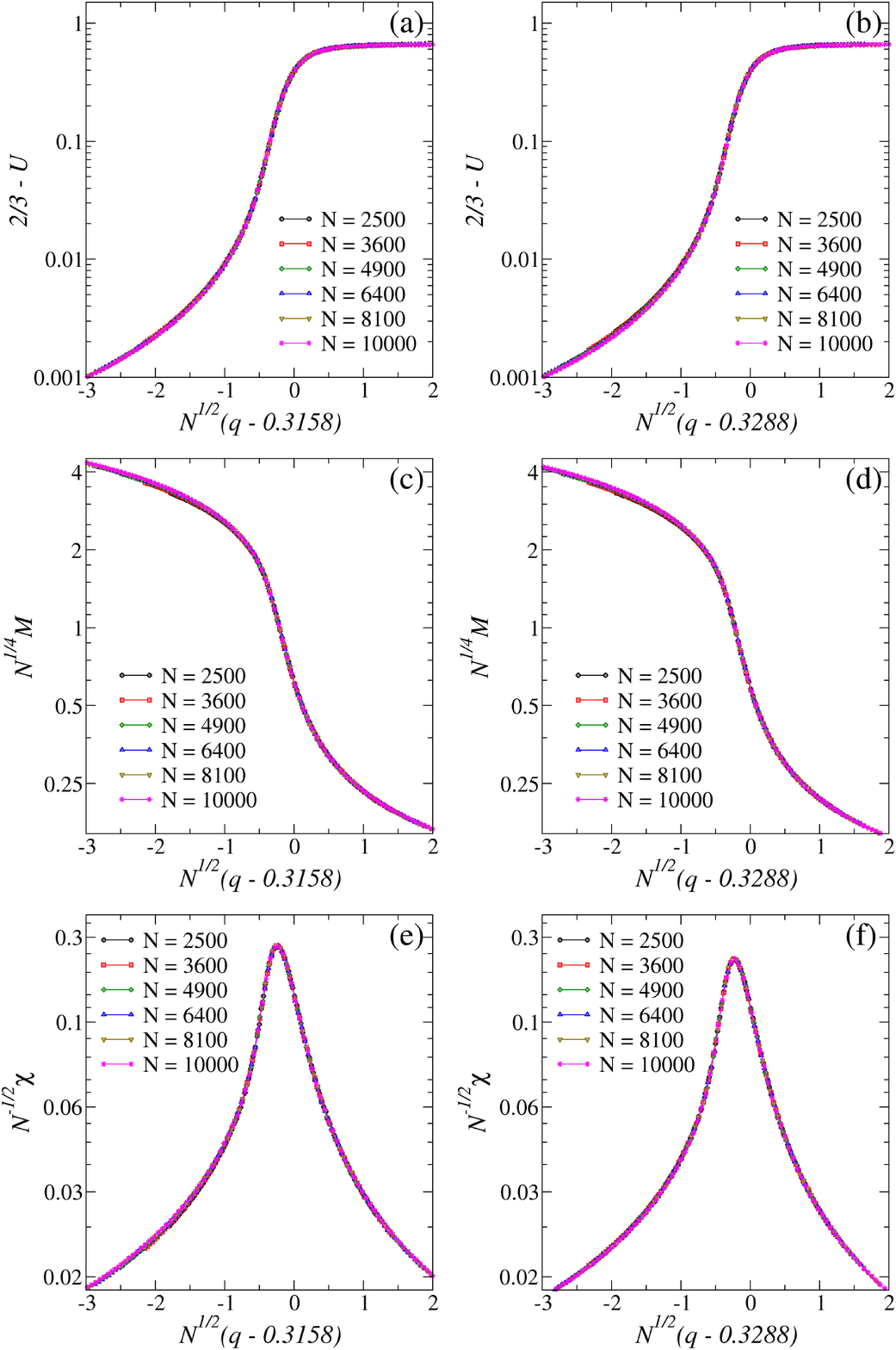} 
\end{center}
\caption{(Color Online) Similar to Fig(\ref{mv-collapsed-results}), but for the BCS model. The BCS model presents a clean critical behavior without logarithmic corrections, and the critical exponents are independent of connectivity $z$. Statistical errors are smaller than symbols.}
\label{bcs-collapsed-results}
\end{figure}

Analogous results of the BCS model showed in Fig.(\ref{bcs-collapsed-results}), present a clean scaling behavior without logarithmic corrections and with leading critical exponents given on Tab.(\ref{criticalbehaviortable}). We summarized the critical noises for $z=4$ and $z=8$ in Tab.(\ref{criticalbehaviorbcstable}). Critical noises should increase when increasing the connectivity $z$ while we can expect a $q=0.5$ saturation value in the same way for the MV model\cite{PhysRevE.96.012304}. We believe this is a consequence of only pairwise interactions between nodes, where the update on every step depends on only one bond of the network. Updated states of a particular node do not depend on all neighbors, avoiding the effects of diverging degree fluctuations.

In summary, leading critical exponent values used in our data collapses, presented in Tab.(\ref{criticalbehaviortable}) are the same for the MV and BCS models, therefore, both models fall in the same universality class. This is expected because both models are in the same universality class for periodic lattices\cite{MJOliveira1992, PhysRevE.94.062317}.

\begin{table}[h]
\begin{center}
\begin{tabularx}{0.5\textwidth}{YY}
\hline
Connectivity & Critical noise \\ 
\hline
$z=4$        & $q_c=0.3158$   \\
$z=8$        & $q_c=0.3288$   \\
\hline                 
\end{tabularx}
\end{center}
\caption{Critical noises of the BCS model as function of connectivity $z$.}
\label{criticalbehaviorbcstable}
\end{table}

\section{Conclusions}

We considered two consensus formation models, namely the MV and BCS models on BA networks. Our numerical results suggest a continuous phase transition in both models, where the critical noises depend on network connectivity. Our numerical results are consistent with both models falling into the same universality class on BA networks, in the same way of periodic lattices. However, the critical behavior of the MV model is subjected to logarithmic corrections on its scaling and its pseudo-exponents are functions of the network connectivity $z$. Additionally, leading critical exponents are the same of the MV model on random Erd\"{o}s-Renyi graphs, placing BA networks and random Erd\"{o}s-Renyi graphs in the same universality class for the MV model. Besides a lacking result of the BCS model on random Erd\"{o}s-Renyi graphs, we can expect that the BCS model should fall in the same universality class of the MV model for random graphs.

\section{Acknowledgments}

We would like to thank CAPES (Coordena\c{c}\~{a}o de Aperfei\c{c}oamento de Pessoal de N\'{\i}vel Superior), CNPq (Conselho Nacional de Desenvolvimento Cient\'{\i}fico e tecnol\'{o}gico), FUNCAP (Funda\c{c}\~{a}o Cearense de Apoio ao Desenvolvimento Cient\'{\i}fico e Tecnol\'{o}gico) and FAPEPI (Funda\c{c}\~{a}o de Amparo a Pesquisa do Estado do Piau\'{\i}) for the financial support. We acknowledge the use of Dietrich Stauffer Computational Physics Lab, Teresina, Brazil, and Laborat\'{o}rio de F\'{\i}sica Te\'{o}rica e Modelagem Computacional - LFTMC, Piripiri, Brazil, where the numerical simulations were performed.

\bibliography{textv1}

\end{document}